\documentclass[ajl]{emulateapj}
\usepackage{graphicx,epsfig,natbib,color}
\usepackage{lscape}
\usepackage{graphicx}
\usepackage{epsfig}
\usepackage{natbib}
\usepackage[colorlinks,citecolor=blue]{hyperref}

\begin{document}
\title{Spectroscopic Diagnostics of Solar Magnetic Flux Ropes Using Iron Forbidden Line}
\author{X. Cheng$^{1,2}$ \& M. D. Ding$^{1,2}$}
\affil{$^1$ School of Astronomy and Space Science, Nanjing University, Nanjing 210093, China}\email{xincheng@nju.edu.cn}
\affil{$^2$ Key Laboratory for Modern Astronomy and Astrophysics (Nanjing University), Ministry of Education, Nanjing 210093, China}

\begin{abstract}
In this Letter, we present Interface Region Imaging Spectrograph \ion{Fe}{21} 1354.08 {\AA} forbidden line emission of two magnetic flux ropes (MFRs) that caused two fast coronal mass ejections with velocities of $\ge$1000 km s$^{-1}$ and strong flares (X1.6 and M6.5) on 2014 September 10 and 2015 June 22, respectively. The EUV images at the 131 {\AA} and 94 {\AA} passbands provided by the Atmospheric Imaging Assembly on board \textit{Solar Dynamics Observatory} reveal that both MFRs initially appear as suspended hot channel-like structures. Interestingly, part of the MFRs is also visible in the \ion{Fe}{21} 1354.08 forbidden line, even prior to the eruption, e.g., for the SOL2014-09-10 event. However, the line emission is very weak and that only appears at a few locations but not the whole structure of the MFRs. This implies that the MFRs could be comprised of different threads with different temperatures and densities, based on the fact that the formation of the \ion{Fe}{21} forbidden line requires a critical temperature ($\sim$11.5 MK) and density. Moreover, the line shows a non-thermal broadening and a blueshift in the early phase. It suggests that magnetic reconnection at that time has initiated; it not only heats the MFR and, at the same time, produces a non-thermal broadening of the Fe XXI line but also produces the poloidal flux, leading to the ascending of the MFRs.
\end{abstract}


\section{Introduction}
Coronal mass ejections (CMEs) are among the most spectacular explosive phenomena in our solar system \citep{chen11_review}. They are able to release a vast amount of high-speed plasma and magnetic flux into the interplanetary space. If the magnetized plasma propagates toward the Earth, it can interact with the magnetosphere and ionosphere, potentially leading to a catastrophic space weather \citep{webb94}. 

Theoretically, the fundamental structure of CMEs is believed to be a magnetic flux rope (MFR), in which the field lines wrap around the central axis and constitute a coherent structure \citep[e.g.,][]{titov99}. When some ideal MHD instabilities such as kink instability \citep{torok04,fan07} and torus instability \citep{kliem06,olmedo10} and/or magnetic reconnection of types such as tether-cutting \citep{moore01} and breakout \citep{antiochos99,chen00,karpen12} occur, the MFR loses its equilibrium \citep{linjun00} and then erupts outward taking the form of a CME \citep{amari14}.

\begin{figure*}
\center {\includegraphics[width=14.5cm]{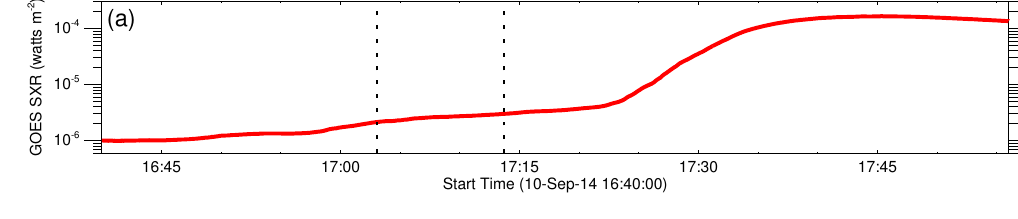}\hspace{0.03\textwidth} \vspace{-0.035\textwidth}}
\center {\includegraphics[width=15cm]{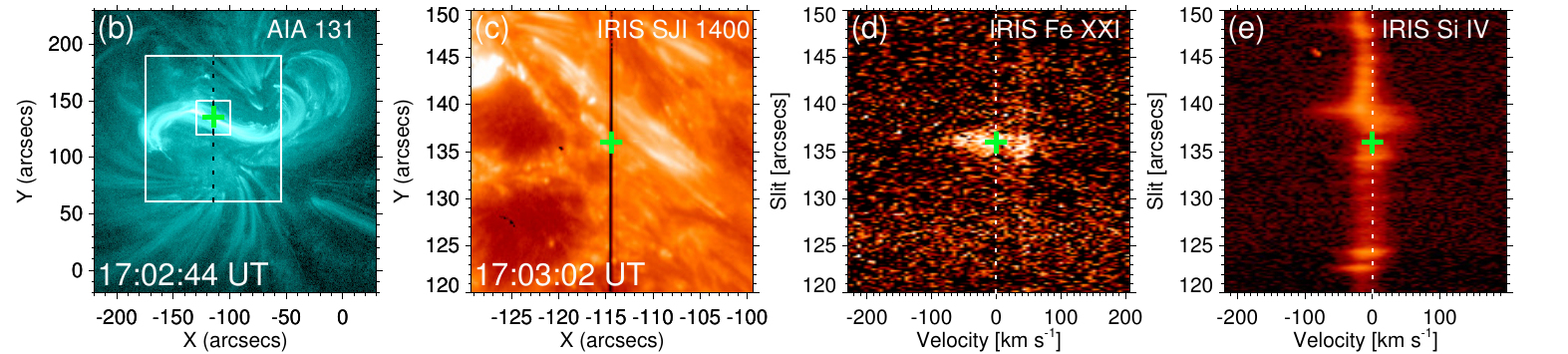}\vspace{-0.02\textwidth}}
\center {\includegraphics[width=15cm]{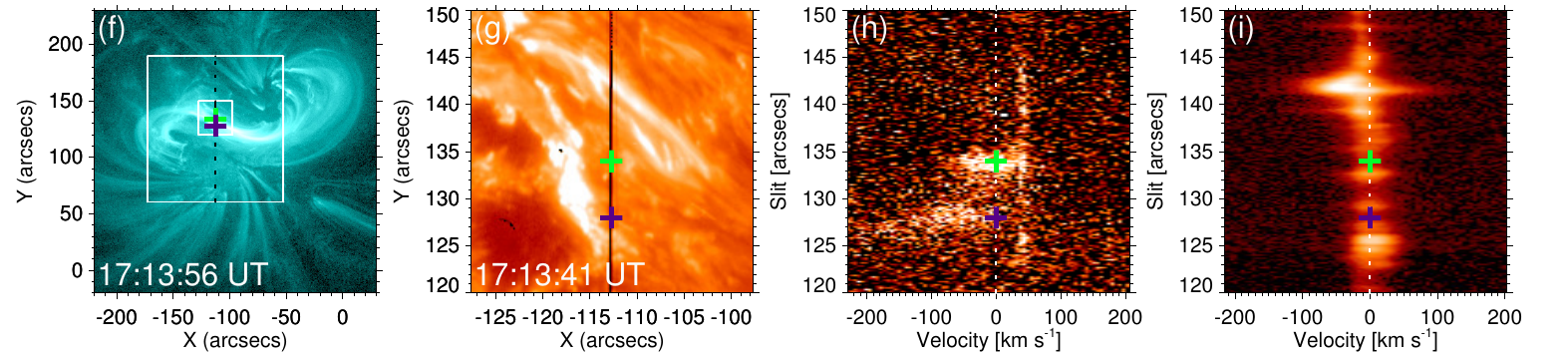}}
\caption{(a) \textit{GOES} SXR 1--8 {\AA} flux of SOL2014-09-10T19:21 flare. (b) \textit{SDO}/AIA 131 {\AA} image showing the morphology of the MFR at 17:02:44 UT (the left vertical dashed line in panel (a)). The bigger white box refers to the whole IRIS FOV, and the smaller one corresponds to that of panels (c), (d), and (e). (c) IRIS 1400 {\AA} SJI. (d) Spectrum of the \ion{Fe}{21} 1354.08 {\AA} line along the slit (the black line in panel (c)). (e) Spectrum of the \ion{Si}{4} 1402.82 {\AA} line along the slit. In panels (b) and (c), the vertical lines represent the position of the slit. In panels (d) and (e), the vertical lines show the theoretical line centers of the \ion{Fe}{21} 1354.08 and \ion{Si}{4} 1402.82 {\AA} lines, respectively. The green plus sign refers to the location of y=136\arcsec. (f)--(i): Similar to panels (b)--(e) but for the time 17:13:56 UT (the right vertical dashed line in panel (a)). The two plus signs point to the locations of y=128\arcsec and y=134\arcsec, respectively.}
\label{f1}
(Animation of this figure is available in the online journal.)
\vspace{0.01\textwidth}
\end{figure*}

Observationally, the MFR is believed to exist in the corona prior to the eruption. In long-decay active regions (ARs), it probably appears as a dark cavity \citep{gibson06_apj,wangym10,dove11,lixing12} or filament channel \citep{vanballegooijen98,low95_apj,gibson06_jgr} as seen in visible, extreme ultraviolet (EUV), and soft X-ray (SXR) passbands. In newly emerged or mature ARs, it likely manifests as a linear feature as seen in forward or reversed sigmoids \citep{canfield99,liur10}. Recently, \citet{zhang12} and \citet{cheng13_driver} discovered a new pattern of manifestation of MFRs, a suspended sigmoidal channel-like structure. This structure is usually not visible in the Atmospheric Imaging Assembly \citep[AIA;][]{lemen12} 131 and 94 {\AA} passbands until tens of minutes before the eruption. After experiencing a slow rise phase, it starts to expand and its morphology is quickly transformed into a semicircular shape \citep{cheng13_driver,lileping13}. At the same time, it drives the formation of the CME and subsequently accelerates the CME with the aid of magnetic reconnection that is supposed to take place near the null point or quasi-separator located between the MFR and flare loops \citep{cheng11_fluxrope,cheng14_tracking,song14_formaion,sun15_nc}. 
 
The most important feature of the channel-like MFR is its high temperature of $\ge$8MK. Utilizing the differential emission measure (DEM) technique, \cite{cheng14_formation} constructed emission measure (EM) maps of the AR 10720 that hosts a well observed hot channel-like MFR. They found that, in the EM map with the temperature range of 8--10 MK, the structure with the strongest emission strikingly resembles the channel-like MFR in the AIA 131 and 94 {\AA} passbands, which confirms that the MFR contains plasma of high temperature. After the eruption of the MFR, the temperature is found to further increase even to 14 MK \citep{cheng11_fluxrope,hannah13,chenbin14}. In some cases, the MFR may erupt but remain confined in the corona. Then, it probably evolves into a ``fire ball" with a temperature of $\sim$10 MK for a relatively long time \citep{song14,tripathi13,cheng14_kink}.

\begin{figure*}
\center {\includegraphics[width=10cm]{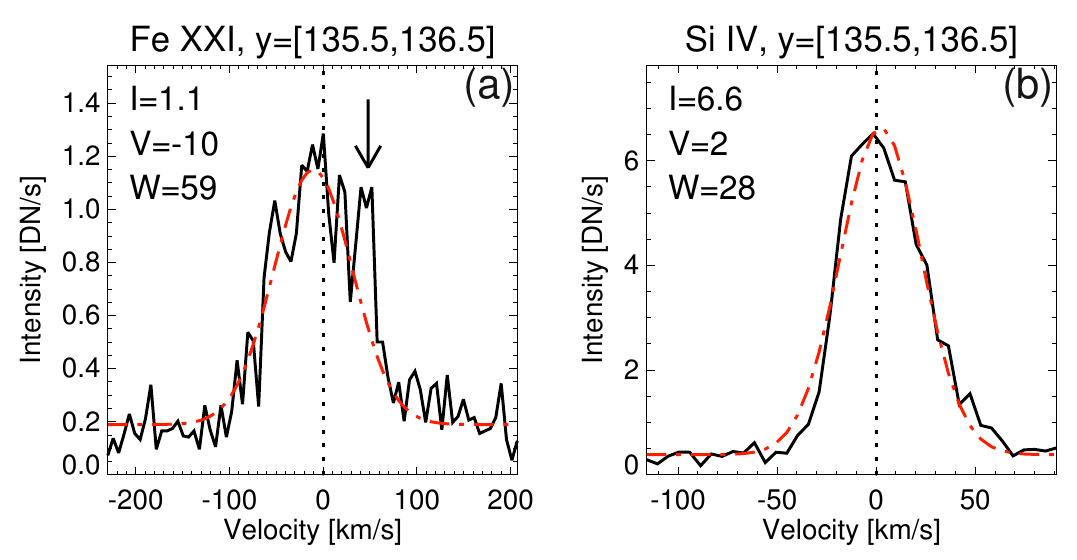}\vspace{-0.03\textwidth}}
\center {\includegraphics[width=10cm]{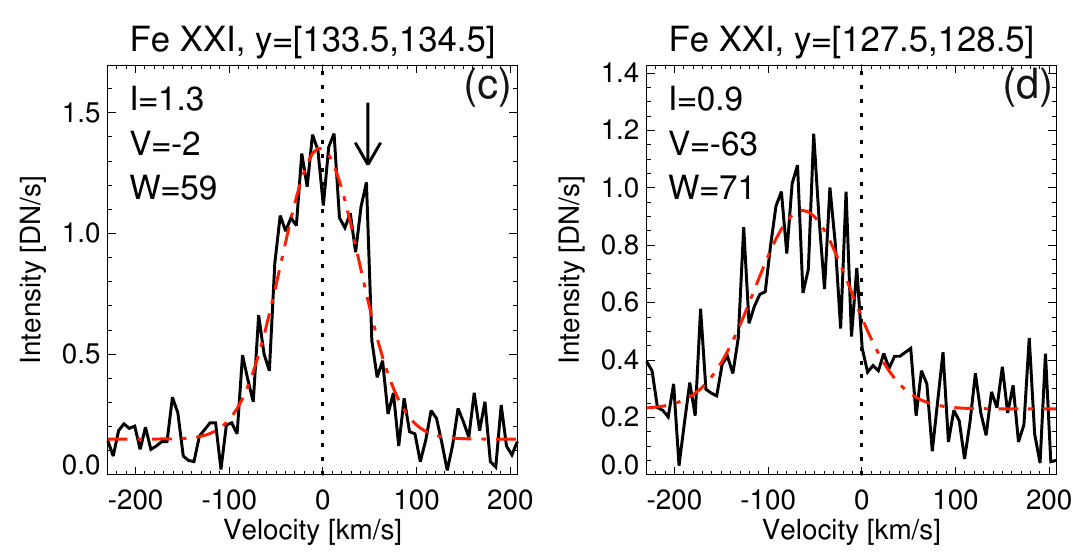}}
\caption{(a) Average profile of the \ion{Fe}{21} 1354.08 {\AA} line over the range of y=[135.5,136.5] at 17:03:02 UT. (b) Average profile of the \ion{Si}{4} 1402.82 {\AA} line for the same location and time as in panel (a). (c) and (d) Average profiles of the \ion{Fe}{21} 1354.08 {\AA} line at 17:13:41 UT over the range of y=[135.5,136.5] and y=[127.5,128.5], respectively. Arrows in panels (a) and (c) point to the \ion{C}{1} 1354.288 {\AA} line. Also shown in all of panels are the single Gaussian fitting to the observed line profiles and the corresponding fitting parameters of peak intensity ($I$), line shift ($V$), and line width ($W$).}
\label{f2}
\vspace{0.01\textwidth}
\end{figure*}

Although it has been verified that the MFR has a very high temperature, some key questions such as what process heats the MFR and where it occurs are still unclear, in particular prior to the eruption. In this Letter, we for the first time use the \ion{Fe}{21} 1354.08 {\AA} forbidden line, provided by the recently launched Interface Region Imaging Spectrograph \citep[IRIS;][]{depontieu14}, to diagnose the dynamics and heating of two MFRs in their early stage. The 1354.08 {\AA} line is the strongest emission line generated by the $^{3}$P$_0$--$^{3}$P$_1$ ground transition of \ion{Fe}{21} with a formation temperature of $\sim$11.5 MK \citep{young15}. It thus provides a promising tool in diagnosing the hot plasma in the MFR. In Section 2, we introduce observational instruments. We then present the results in Section 3, followed by a summary and discussion in Section 4.

\section{Instruments}
The observational data are mainly from \textit{Solar Dynamics Observatory} \citep[\textit{SDO};][]{pesnell12} and IRIS. The AIA on board \textit{SDO} provides EUV images of the solar corona with a temporal cadence of 12 s and spatial resolution of 1.2\arcsec. Here, we only use the 131~{\AA}, 94 {\AA}, and 193 {\AA} passbands because of their high sensitivity to the high temperature plasma ($\ge$7 MK). IRIS provides spectra of the chromosphere, transition region, and corona with a spatial resolution of 0.33--0.4\arcsec, temporal cadence of $\sim$2 s, and spectral resolution of $\sim$1 km s$^{-1}$ \citep{depontieu14}. The spectral lines we used include the \ion{Si}{4} 1394/1403 {\AA} lines formed in the transition region with a temperature of $\sim$0.06 MK and the \ion{Fe}{21} 1354.08 {\AA} line formed in the corona with a temperature of $\sim$11.5 MK. Simultaneously, we also use the slit-jaw images (SJIs) at the 1400 {\AA} passband. The absolute wavelengths of the \ion{Fe}{21} and \ion{Si}{4} lines are calibrated by assuming a zero velocity of the nearby \ion{O}{1} 1355.5977 {\AA} and \ion{Fe}{2} 1405.608 {\AA} lines, respectively \citep[also see;][]{tian15}. Moreover, the \textit{GOES} provides the SXR 1--8 {\AA} flux of the associated flares.

\section{Results}
In order to search for the visibility of the hot channel-like MFRs in the \ion{Fe}{21} line, we inspect all of the flares\footnote{http://iris.lmsal.com/data.html} that have been observed by IRIS and are associated with hot channel-like MFRs in the AIA~94 {\AA} or 131 {\AA} images. Note that we only focus on the early phase of the flare during which the MFRs just come into view and can be fully covered by the field-of-view (FOV) of IRIS (175\arcsec$\times$175\arcsec). Finally, we find two events, SOL2014-09-10 and SOL2015-06-22, both of which present clear \ion{Fe}{21} line emission at the locations of the MFRs. For the first event, the slit of IRIS is set as sit-and-stare mode and positioned to be perpendicular to the axis of the MFR; while for the second one, the slit is set as 16-step raster mode and almost parallel with the axis of the MFR.

\begin{figure*}
\center {\includegraphics[width=14.5cm]{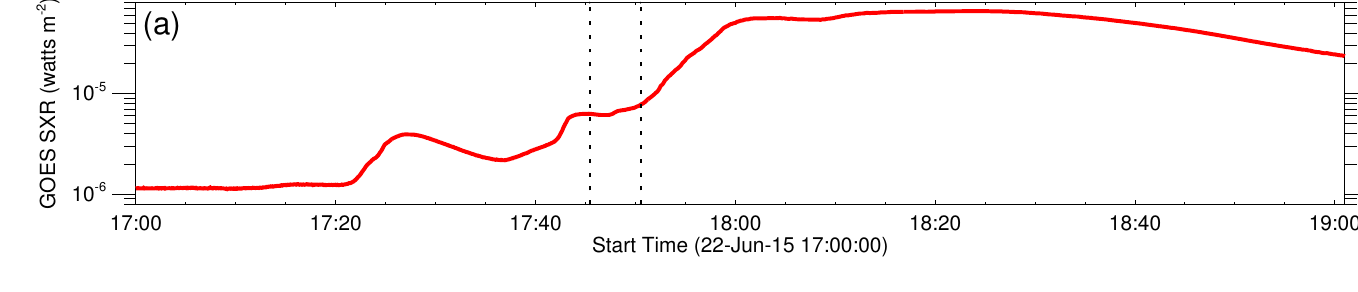}\hspace{0.03\textwidth} \vspace{-0.035\textwidth}}
\center {\includegraphics[width=15cm]{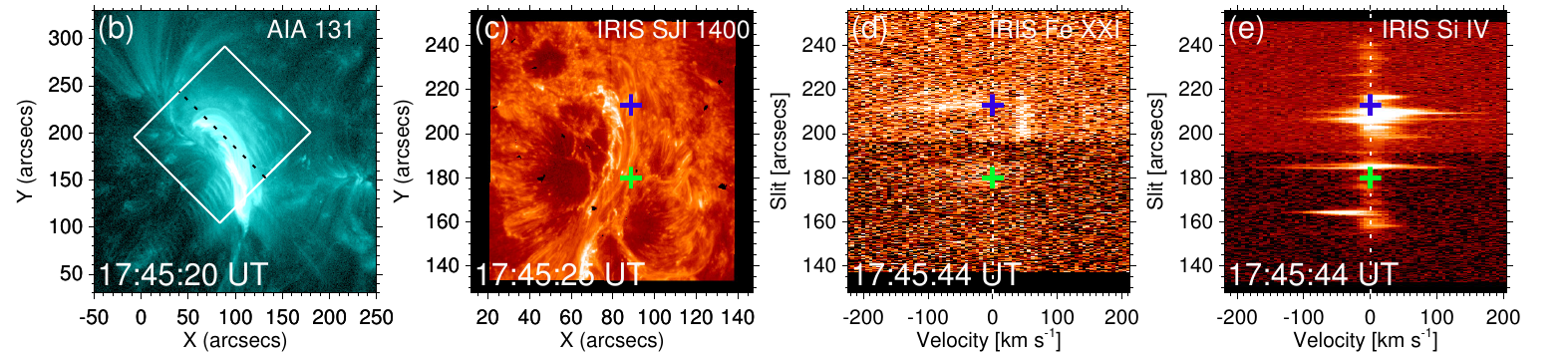}\vspace{-0.02\textwidth}}
\center {\includegraphics[width=15cm]{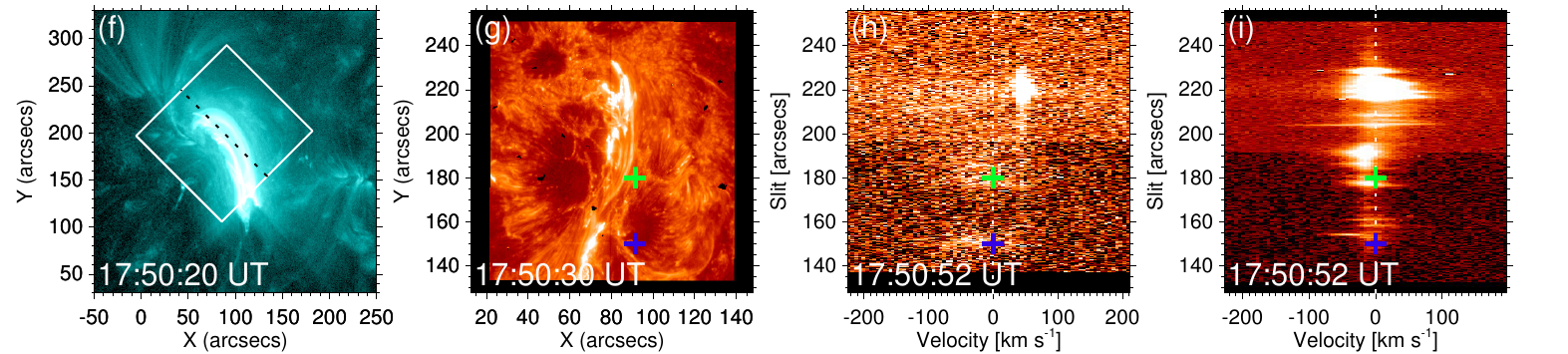}}
\caption{(a) \textit{GOES} SXR 1--8 {\AA} flux of SOL2015-06-22T17:39 flare. (b) \textit{SDO}/AIA 131 {\AA} image showing the morphology of the MFR at 17:45:20 UT (the left vertical line in panel (a)). The white box refers to the FOV of IRIS, the oblique dash line corresponds to the position of the slit. (c) IRIS 1400 {\AA} SJI. (d) and (e) Spectrum of the \ion{Fe}{21} 1354.08 {\AA} line and the \ion{Si}{4} 1402.82 {\AA} line along the slit. The vertical lines show the theoretical line centers of the \ion{Fe}{21} 1354.08 and \ion{Si}{4} 1402.82 {\AA}, respectively. The two plus signs point to the locations of y=180\arcsec and y=213\arcsec, respectively. (f)--(i) Similar to panels (b)--(e) but for the time 17:50:20 UT (the right vertical line in panel (a)). The two plus signs refer to the locations of y=180\arcsec and y=150\arcsec, respectively.}
\label{f3}
(Animation of this figure is available in the online journal.)
\vspace{0.01\textwidth}
\end{figure*}

\subsection{The 2014 September 10 Event}
The SOL2014-09-10 hot channel-like MFR originates in NOAA AR 12158 that is located at N11E05. The eruption of the MFR produces a CME with a projected speed of $\sim$1260 km s$^{-1}$ and an X1.6 flare. In our previous work, we found that the twist field in the MFR has been well build up before 17:00 UT, which is mainly attributed to the reconnection between two groups of sheared arcades in the lower atmosphere as evidenced by the large blueshift and strong non-thermal broadening of the IRIS \ion{Si}{4}, \ion{C}{2}, and \ion{Mg}{2} lines \citep[see][]{cheng15_iris}. According to the \textit{GOES} record, the flare started at 17:21 UT and peaked at 17:45 UT (Figure \ref{f1}a). However, actually, some sigmoidal threads in the MFR have started to brighten and rise up since as early as $\sim$17:00 UT (Figure \ref{f1}b and \ref{f1}f). Considering the projection effect, such a rise displays a movement toward the south. At the same time, the SXR emission increases slightly. The temperature of the threads is estimated to be $\ge$7 MK because they are only visible in the AIA 131 {\AA} and 94 {\AA} passbands but not in other relatively cool AIA passbands.

The \ion{Fe}{21} 1354 {\AA} line is only seen in the early phase at the SOL2014-09-10 hot channel-like MFR (Figure \ref{f1} and attached online movie). At 17:00 UT, there appears a weak emission of the line at y=[134,137]. With the time elapsing, the intensity gradually increases. At 17:03 UT, one can see an apparent \ion{Fe}{21} line emission (Figure \ref{f1}d). Figure \ref{f2}a displays the average line profile over the range of y=[135.5,136.5]. It is seen that the peak intensity is only $\sim$1 DN s$^{-1}$. Applying the single Gaussian fitting, we get a blueward line shift (an upward Doppler velocity) of $\sim$10 km s$^{-1}$ and a line width of $\sim$59 km s$^{-1}$, respectively. Using the routine iris\_non-thermalwidth.pro in the SSW package, we derive a non-thermal velocity of 18 km s$^{-1}$. In order to check the altitude of the hot plasma in the solar atmosphere, we also inspect the IRIS 1400 {\AA} SJIs (Figure \ref{f1}c) and the \ion{Si}{4} 1403 {\AA} spectrum (Figure \ref{f1}e) that are thought to reflect the heating status of the transition region and the upper chromosphere. It is found that the 1400 {\AA} intensity and the \ion{Si}{4} line profile (Figure \ref{f2}b) at the projected location of the MFR are very similar to that at the quiet region. It implies that the \ion{Fe}{21} 1354.08 {\AA} line emission is most likely from the hot plasma in the corona rather than somewhere below.

As the MFR rises up, the \ion{Fe}{21} line emission starts to appear at the second location from $\sim$17:13 UT (y=[127.5,128.5]; Figure \ref{f1}h). At 17:15 UT, the line emission reaches its maximum. Compared with the line profile at y=[135.5,136.5] (Figure \ref{f2}c), the average profile at y=[127.5,128.5] (Figure \ref{f2}d) shows a significant shift toward the blue wing (--63 km s$^{-1}$). Moreover, the line width increases to 71 km s$^{-1}$, which reduces to a non-thermal velocity of 43 km s$^{-1}$. Similarly, we also inspect the 1400 {\AA} intensity and \ion{Si}{4} line at the MFR location that do not show any difference from that in the quiet region.

\subsection{The 2015 June 22 Event}
The SOL2015-06-22 MFR originates in NOAA AR 12371 that is located at N13W14. Its eruption causes a CME with a projected speed of $\sim$1000 km s$^{-1}$ \citep{liuying2015}. The associated flare (M6.5 class) started at 17:39 UT and peaked at 18:23 UT (Figure \ref{f3}a). Note that before the flare, there occurs another C3.9 flare starting at $\sim$17:20 UT. However, through carefully inspecting its source region, it is found that the C3.9 flare is actually from AR 12367 (S18W78) but not from AR 12371 (N13W14).

It is found that a loop-like structure along the polarity inversion line starts to brighten prior to the flare (attached movie of Figure \ref{f3}). This structure mostly appears in the AIA 131 {\AA}, 94 {\AA}, and 193 {\AA} passbands but not in the other passbands, showing a temperature of $\ge$7 MK. Moreover, in the AIA 304 {\AA} and 171 {\AA} passbands, there appear three small filaments that are cospatial with the hot loop-like structure. It implies that the hot structure and the filaments may belong to the same magnetic system. Different from the SOL2014-09-10 MFR, the slow rise of the SOL2015-06-22 one occurs in the period of $\sim$17:35--17:51 UT, which coincides with the early phase of the flare, i.e., from the start of the flare (17:39 UT) to the start of the fast rise of the SXR flux ($\sim$17:51 UT). In the early phase of the flare, the bright structure first rises detaching from the sheared loops underneath and then quickly moves upward (Figure \ref{f3}b and \ref{f3}f). When ascending to a certain height, this bright structure displays a writhed morphology ($\sim$18:00 UT), which is, however, quickly transformed into a semicircular shape. Such properties confirm that the erupting bright structure is indeed a hot channel-like MFR \citep[e.g.,][]{zhang12,cheng13_driver,cheng14_tracking}. 

\begin{figure*}
\center {\includegraphics[width=15cm]{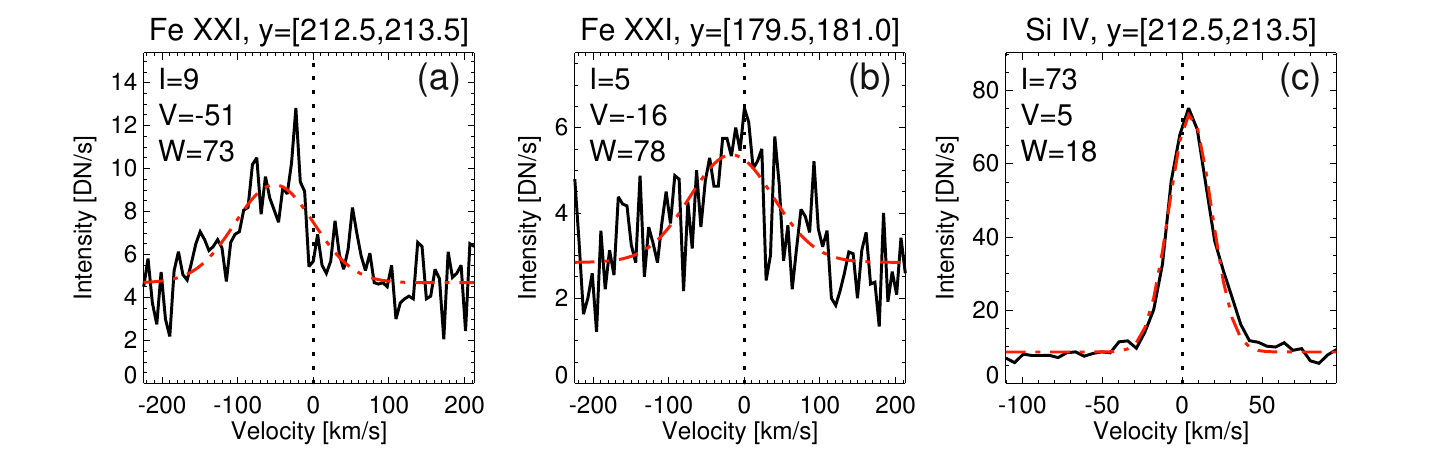}\vspace{-0.025\textwidth}}
\center {\includegraphics[width=15cm]{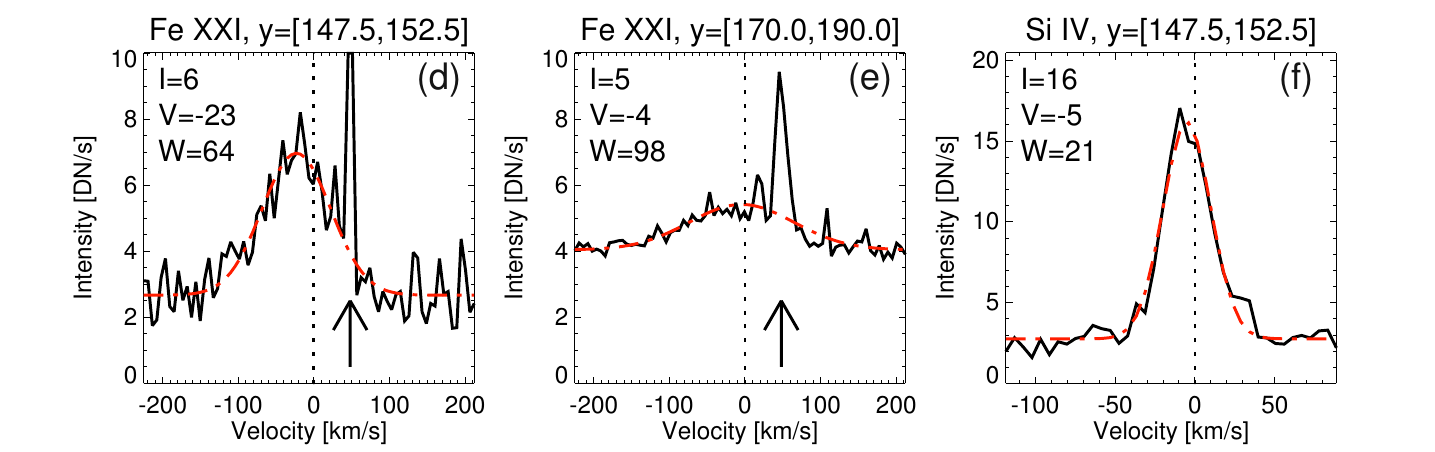}\vspace{0.0\textwidth}}
\caption{(a) and (b) Average profiles of the \ion{Fe}{21} 1354.08 {\AA} line over the range of y=[212.5,213.5] and y=[179.5,181.0] at 17:45:44 UT. (c) Average profile of the \ion{Si}{4} 1402.82 {\AA} line for the same location and time as in panel (a). (d) and (e) Average profiles of the Fe XXI 1354.08 {\AA} line over the range of y=[147.5,152.5] and y=[170.0,190.0] at 17:50:52 UT. The two arrows point to the \ion{C}{1} 1354.288 {\AA} line. (f) Same as (c) but for 17:50:52 UT. In all of panels, the dot-dashed curves refer to the single Gaussian fitting to the observed line profiles.}
\vspace{0.02\textwidth}
\label{f4}
\end{figure*}

The SOL2015-06-22 MFR is also observed by the IRIS in the \ion{Fe}{21} line. It can be found from Figure \ref{f3}d that, at $\sim$17:45:44 UT (the raster step=9), the \ion{Fe}{21} line emission appears at two positions ([x,y]=[87, 213] and [87,180]). In the range of y=[212.5,213.5], the peak of the average profile is 9 DN s$^{-1}$ (Figure \ref{f4}a). Applying the single Gaussian fitting to the line profile, the blueward line shift (upward Doppler velocity) and non-thermal velocity are found to be about 51 and 45 km s$^{-1}$, respectively. At y=[179.5,181.0], the \ion{Fe}{21} line is relatively weak (5 DN s$^{-1}$; Figure \ref{f4}b). The corresponding Doppler and non-thermal velocities are $\sim$16 and 53 km s$^{-1}$, respectively.

The strongest emission of the \ion{Fe}{21} 1354.08 {\AA} line appears at the right part of the MFR, i.e., the position near $\sim$y=160. Since the MFR was scanned by the slit (with a 16-step mode) that is roughly parallel with the axis of the MFR, we can check the spatial variation of the \ion{Fe}{21} line emission along the MFR and its temporal behavior. It is seen that the \ion{Fe}{21} line appears in sequence from the leg (y=150) to the top (y=190) of the MFR though the absolute intensity tends to decrease. This behavior starts at 17:43:44 UT and lasts till 17:50 UT, after which the emission of the \ion{Fe}{21} line mostly comes from the post-flare loops and thus becomes very strong. More details on the spatial variation and evolution of the line can be seen from the attached movie of Figure \ref{f3}.

Figure \ref{f3}h shows the spectrum of the \ion{Fe}{21} line at 17:50:52 UT (the raster step of 11). At the range of y=[147.5,152.5], it exhibits a blueshited velocity of 23 km s$^{-1}$ and a non-thermal velocity of $\sim$30 km s$^{-1}$ (Figure \ref{f4}d). While at the range of y=[170.0,190,0], the main feature of the line is a significant non-thermal broadening that is as large as 80 km s$^{-1}$ (Figure \ref{f4}e). Finally, we also inspect the spectrum (Figure \ref{f3}i) and average profiles (Figure \ref{f4}c and \ref{f4}f) of the \ion{Si}{4} 1403 {\AA} line at the MFR that again show no obvious difference from that in the quiet region, similarly to the SOL2015-06-22 MFR.

\section{Summary and Discussion}
The MFR is thought to play an important role in the early dynamics of CMEs and flares. Previous studies revealed that it probably exists prior to the eruption and has a temperature as high as $\sim$10 MK. In this Letter, we for the first time use the \ion{Fe}{21} 1354.08 {\AA} forbidden line to diagnose the thermo dynamics of the MFRs. The \ion{Fe}{21} line has often been seen at post-flare loops and flare ribbons \citep[e.g.,][]{tian14_reconnection,liying15,young15}. However, we find that this line also appears at the MFRs in the early phase of CMEs and flares, even prior to the eruption, e.g., for the SOL2014-09-10 event, although the line emission is very weak. In the pre-flare (early) phase, the counts of the line are only $\sim$1 (10) DN s$^{-1}$, far less than the values of $\sim$100--1000 DN s$^{-1}$ at the flare loops and ribbons \citep[e.g.,][]{graham15,tian15,lidong15,liying15,young15,liting15_slipping}.

It is found that the \ion{Fe}{21} 1354.08 {\AA} line only appears at a few locations of the MFRs. For the SOL2014-09-10 MFR, the IRIS slit is positioned crossing the MFR axis with a section of at least $\sim$30\arcsec~in size. However, only at a small part of the crossing section with a size less than 10\arcsec, the \ion{Fe}{21} line emission can be seen. While for the SOL2015-06-22 MFR, most part of the structure is overlapped by the slit; but similarly, only part of it presents the \ion{Fe}{21} line emission. Considering that the \ion{Fe}{21} 1354 {\AA} line is a forbidden line whose formation requires a critical temperature (11.5 MK) and density \citep{young15}, our results show that, although the MFR could be heated to a high temperature, only some hot threads within are able to produce detectable line emission. This implies that the MFR could be comprised of multiple threads with quite different temperatures and densities. 

Moreover, the \ion{Fe}{21} 1354.08 {\AA} line at the MFR displays a significant non-thermal broadening and an apparent blueshift in the early phase. For the SOL2014-09-10 MFR, such a blueshift even appears prior to the eruption, varying in a range of 10--60 km s$^{-1}$ for the part covered by the slit. Correspondingly, the non-thermal velocity varies in a range of 18--43 km s$^{-1}$. The results suggest that magnetic reconnection, either inside or surrounding the MFR \citep{cheng13_double,harra13}, has started prior to the eruption with, though, a relatively low efficiency. Such a reconnection can heat the MFR and, at the same time, produces a non-thermal broadening of the Fe XXI line. On the other hand, the reconnection can also provide additional poloidal flux to the MFRs that results in an upward Lorentz self-force to drive the motion of the MFRs \citep[also see;][]{syntelis16}. With the flare beginning, the reconnection rate further increases, leading to a more apparent non-thermal broadening of the Fe XXI line, just as what revealed in the SOL2015-06-22T event.

\acknowledgements We are grateful to the referee for his/her valuable comments that improved the manuscript. We also thank Jie Zhang and Hui Tian for their useful discussions. \textit{SDO} is a mission of NASAs Living With a Star Program. IRIS is a NASA small explorer mission developed and operated by LMSAL with mission operations executed at NASA Ames Research center and major contributions to downlink communications funded by the Norwegian Space Center (NSC, Norway) through an ESA PRODEX contract. This work is supported by NSFC under grants 11303016, 11373023, and NKBRSF under grant 2014CB744203.


\end{document}